\documentclass[%
 reprint,
 amsmath,amssymb,
 aps,
nofootinbib]{revtex4-1}
\usepackage{subcaption}
\usepackage{color}
\usepackage{graphicx}
\usepackage{dcolumn}
\usepackage{bm}
\usepackage{physics}
\usepackage{blindtext}
\usepackage{natbib}
\usepackage{float}
\usepackage{graphicx,wrapfig}
 \newcommand*\diff{\mathop{}\!\mathrm{d}}
 
\begin{document}


\title{Quantum Mechanical Treatment of Two-Level Atoms Coupled to Continuum with an Ultraviolet Cutoff}

\author{Fatih Din\c{c}}
 \email{fatih.dinc@boun.edu.tr}
 \affiliation{Electrical \& Electronics Engineering Department, Bo\u{g}azi\c{c}i University, Istanbul, 34342, Turkey}

 \author{\.{I}lke Ercan}
  \email{ilke.ercan@boun.edu.tr}
 \affiliation{Electrical \& Electronics Engineering Department, Bo\u{g}azi\c{c}i University, Istanbul, 34342, Turkey}

\date{\today}

\begin{abstract}
In this paper, we provide a rigorous quantum mechanical derivation for the coherent photon transport characteristics of a two-level atom coupled to a waveguide without linearizing the coupling coefficient between the light and the atom. We propose a novel single frequency sampling method utilizing a UV-cutoff that allows us to treat the singularities in real space scattering potential despite the non-convergence property.  We also study the conditions under which the linearization of the coupling coefficient is an accurate assumption and find the resulting spontaneous emission and transport characteristics taking the radiative and non-radiative decay rates into account. This allows us to confirm and expand on the findings of the existing literature while obtaining the dynamic electronic polarizability for the two-level atom confined to a 1-D waveguide while using an interaction Hamiltonian with rotating-wave approximation.

\end{abstract}


\maketitle

\section{Introduction}
The potential realizations of light-matter interactions pave the wave for novel applications in emerging technologies from quantum information processing to single atom switches. The studies on two-level atom and photon interactions lay out the theoretical foundations of such systems. However, the definition of a two-level system is a broad one and the current formalism involves certain assumptions which, in some cases, restrict the scope of interactions we can analyse.  Under the existing formalism, the explanation of  transport properties resulting from the spontaneous emission involves assuming that the coupling between the two-level atom and photon is weak. The accuracy of this assumptions is currently underexplored and can be questioned for certain coupling regimes.  \par 

The foundations laid out by the existing formalism and experimental opportunities for the two-level atom and photon interactions has allowed treating complex systems (e.g. Ref. \cite{3-level,bos,bos-2,bos-3,bos-4,bos-5,bos-6,bos-7,bos-8,bos-9,bos-10,
ilkehoca-1,ilkehoca-2,ilkehoca-3}). The approach proposed in Ref.s \cite{fan,Law}  presents a treatment of two-level systems in real space, while the Dicke--Hamiltonian \cite{dicke-Hamiltonian} is transformed into the real space and used to predict a mirror behavior of two-level atoms for certain wavelength of light. However, this predominant approach involves the assumption that the coupling coefficient $V_k$  is linearized, which leads to an approximate solution in the weak coupling regime.\par 
 It is also shown classically in Ref. \cite{photonics} that solving the wave equation for the electric field for a delta-shape dynamic electronic polarizability $a(w)$ leads to certain transmission and reflection coefficients which are dependent on the polarizability $a(w)$. An equivalent analogy can be drawn between the results of the Ref.s \cite{fan,photonics}, from which the dynamic electronic polarizability of the two-level atom can be deduced. However, the resulting dynamic electronic polarizability $a(w)$ does not agree with the Ref.  \cite{polarizability}. This stems from the fact that, the process considered in Ref. \cite{fan} requires not an approximate but an exact treatment of the coupling between light and the two-level atom, without ignoring the relation between the coupling coefficient and the energy of the incident photon. \par
 In this paper, we address this need by proposing a rigorous quantum mechanical derivation which allows treatment of two-level atoms coupled to a waveguide without linearizing the coupling coefficient between the light and the atom. As a proof of concept, we find the dynamic electronic polarizability of the TLS through a comparison with the inhomogenous wave equation. \par 
The first look into the exact treatment of the TLS-photon interactions reveal that there is an inherent divergency in the coupling coefficient $V(k)$. As one tries to find the scattering potential $V(x)$ corresponding to this coupling coefficients,  a divergent Fourier transform calls for a need of UV-cutoff frequency $w_{max}=\hbar k_{max}$. Nonetheless, this is not the unique case that nature emposes a UV-cutoff \cite{cutoff,cutoff-2}. In many practical cases, this cutoff can be taken as the frequency after which approximations fail. In the case of TLS coupled to a waveguide, it is natural to impose the cutoff frequency $w_c$ of the waveguide as the maxium frequency $w_{max}$ attainable by the photon. \par
The organization of the paper is as follows: In the next section, we lay out the foundations of our derivation by first outlining  the existing literature and incorporating a general scattering potential. We then propose a sampling method utilizing UV-cutoff that involves expanding our approach on the treatment of non-convergent scattering potentials in real space. In Sec. III, we discuss further characteristics of this system based on the foundations we propose such as the scattering coefficients, S-matrix and excitation probability of the two-level atom, and provide a proof of concept. We also discuss the conditions under which the linearization assumption is valid. In the final section, we conclude with final remarks and comments on future work.

\section{Foundations}
In this section, we lay out the theoretical basis of our approach by first outlining the preliminaries based on the existing literature and then proposing a novel single freuquency-sampling method utilizing an ultraviolet cutoff that allows us to treat analytic functions with diverging position space components, such as coupling coefficient between the two-level system (TLS) and light. The configuration we study in this paper is depicted in Fig. 1; where  the incident photon direction is shown with corresponding arrows relative to the TLS. 
\subsection{Preliminaries}

 The departure point of our derivation is the real space Hamiltonian given in Ref. \cite{fan} modified for a general scattering potential $V(x)$. We carry out calculations with minimum restrictions for the shape of the scattering potential $V(x)$, which is a consequence of the coupling coefficient $V(k) = \sqrt{\frac{L}{2\pi}} V_k = \sqrt{\frac{ \hbar w_k}{4 \pi \epsilon_0 A}} \bar d e^{i (\theta - kx_0)}$ between the light and the TLS  \cite{coupling}. Here $\bar d$ is the average dipole moment of the atom, $\theta$ is an arbitrary phase, $A$ is the waveguide area and $x_0$ is the position of the two-level atom, where the dependency of phase upon position allows one to study atoms in a linear chain \cite{Law}.

\begin{figure}
\centering
\includegraphics[width=6cm]{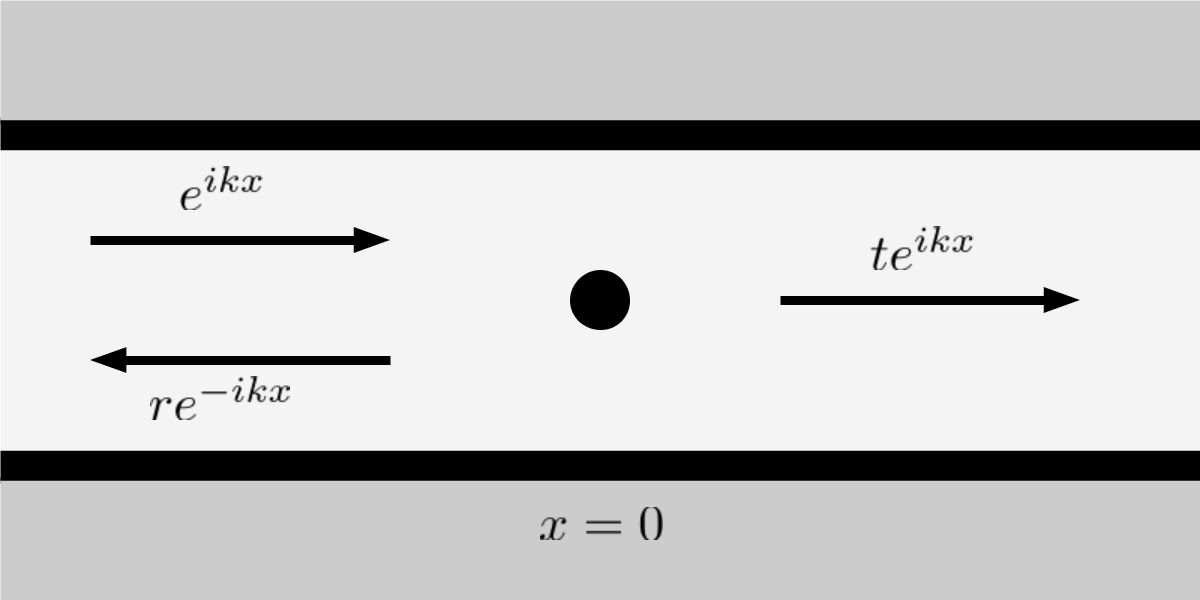} 
\caption{The TLS inside a waveguide upon constant coherent photon incident from left.} \label{figmain}
\end{figure}

The real space Dicke-Hamiltonian for the photon-two-level atom system described in Fig. \ref{figmain} is given as: 

\scriptsize
\begin{equation}  \label{eq:dicke-Hamiltonian}
	\begin{split}
		\hat H &= -i \hbar v_g  \int_{-\infty}^\infty \diff x \, \left(   C_R^\dag(x) \frac{\partial}{\partial x} C_R(x)-  \int_{-\infty}^\infty \diff x \,   C_L^\dag(x) \frac{\partial}{\partial x} C_L(x) \right)\\ 
				&+   \int_{-\infty}^{\infty}\diff x ( V(x) (C_R^\dag(x) + C_L^\dag(x)) S_-  +V^*(x) (C_R(x) + C_L(x))S_+ ) 	\\	
				&+ E_e a_e^\dag a_e \: + E_g \:a_g^\dag a_g  
	\end{split}
\end{equation} \normalsize  
where $C_{R/L}^\dag(x)$ ($C_{R/L}(x)$) are right/left moving particle creation (annihilation) operators,  $E_e$  and $E_g$ represent the energy of the excited and ground state of the atom, respectively, with energy difference $\Omega=E_e-E_g$, $S_-=a^\dag_g a_e$ ($S_+=a^\dag_e a_g$) is the atomic de-excitation (excitation) operators -- i.e. $a_{e/g}=\ket{0}\bra{e/g}$ ($a^\dag_{e/g}=\ket{e/g}\bra{0}$) is the annihilation (creation) operator for the excited/ground state of the atom and $V(x)$ is the scattering potential in real space between the photon and the two-level atom, related to the coupling coefficient in continuous k-space, $V(k)$, as: 
\begin{align}
V(x) &=\frac{1}{\sqrt{2\pi}} \int_{-k_{max}}^{k_{max}} \diff  k   V(k) e^{i kx}
\end{align}
where we have introduced an ultraviolet cutoff $k_{max}$ in order for the Fourier transform to converge. The value of the cutoff is not of importance as the final results will be independent. For the most general case, we can assume that $\hbar k_{max}\sim w_c$, where $w_c$ is the cutoff frequency of the waveguide. For the rest of the article, we shall analyze the TLS from its comoving reference frame to avoid any discussion regarding the invariance of the cutoff frequency.

A possible scattering eigenstate for this Hamiltonian is given by the equation:  

\scriptsize
\begin{equation}  \label{eq:eigenstate}
	\ket{E_k}=\int_{-\infty}^{\infty} \diff x \: \Bigg( \phi_R (x) C_R^\dag(x) + \phi_L (x) C_L^\dag(x) \Bigg) \ket{0,-} + e_k  \ket{0,+}
\end{equation} \normalsize  
where we denote $e_k$ as the probability coefficient\footnote{We note that this eigenstate is not normalized, however this poses no problem for our treatment.} for the excitation of the atom, and $\phi_R(x)$ and  $\phi_L(x)$ are the complex amplitudes of the right and left moving photons, respectively.

For a photon incident far form the left, $\phi_R(x)$ and  $\phi_L(x)$ are taken as ansatz:  
\begin{equation} 
	\phi_R(x)= u_R(x) e^{ikx} \quad \text{and} \quad \phi_L(x)= u_L(x) e^{-ikx}
\end{equation}  
where the complex amplitudes take the form shown in Fig. \ref{figmain}.

\begin{figure}
\centering
\includegraphics[width=9cm]{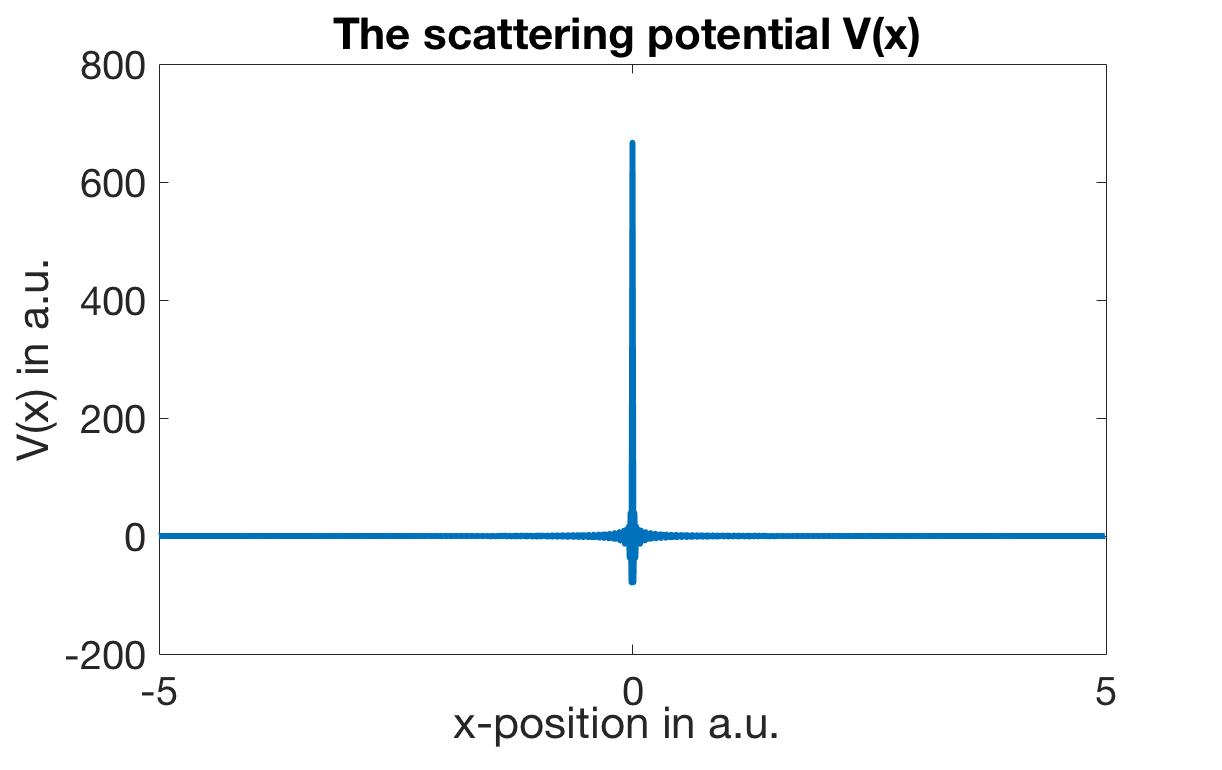} 
\caption{The scattering potential $V(x)$ computed with an ultraviolet cutoff $k_{max}$.} \label{figpot}
\end{figure}

Working in Schr\"odinger picture, eigenvalue equation $\hat H \ket{E_k} = E_k \ket{E_k}$ should be satisfied. The states $\ket{E_k}$ are said the be energy-eigenstates and diagonalize the Hamiltonian. After straightforward algebra, we find out that the energy-eigenstates should satisfy the following conditions:
\begin{subequations} \label{eq:boundary}
\begin{align} 
(\Omega-E_k) e_k  + \int_{-\infty}^{\infty}\diff x V^*(x) \left( \phi_R (x) + \phi_L (x) \right) &= 0 \label{eq:integral} \\ 
- i \hbar v_g \frac{\partial u_R (x)}{\partial x} e^{ikx} + e_k V(x) &= 0, \label{eq:boundary2} \\
  i \hbar v_g \frac{\partial u_L (x)}{\partial x} e^{-ikx} + e_k V(x) &= 0.\label{eq:boundary3}
\end{align}
\end{subequations} 
Here, the shape of the scattering potential $V(x)$ is of significance in order to continue the derivation. With the introduction of ultraviolett cutoff, the Fourier transform $V(x)$ of the coupling potential $V(k)$ converges and is sharply peaked at the origin. A sketch of this potential is given in Fig. \ref{figpot}.

To evaluate the integral in Eqn. (\ref{eq:integral}), we express the wave function in real space as a function of the complex amplitudes of the right and left moving photons:
\begin{equation}
\psi(x) = \phi_R(x) + \phi_L(x)
\end{equation}
then the integral becomes: 

\scriptsize
\begin{align*} 
\int_{-\infty}^{\infty}\diff x V^*(x) \psi(x)&=  \frac{1}{\sqrt{2\pi}} \int_{-\infty}^\infty \int_{-k_{max}}^{k_{max}} \diff x \diff k' V^*(k') e^{-ik'x} \psi(x) \\
&=  \int_{-k_{max}}^{k_{max}}  \diff k' V^*(k') \left( \int_{-\infty}^\infty \diff x \frac{1}{\sqrt{2\pi}} e^{-ik'x} \psi(x) \right) \\
\end{align*}  \normalsize
 where in the last step we recognize the Fourier component of the real space wave-function. With the motivation that we are only interested in its value around $x=0$, the Fourier component $\bar \psi(k')$ can be approximated  as: 
 \begin{equation}
 \begin{split}
 \bar \psi(k') &= \left( \int_{-\infty}^\infty \diff x \frac{1}{\sqrt{2\pi}} e^{-ik'x} \psi(x) \right) \\ &\simeq \sqrt{2 \pi} \left( \frac{1+t}{2} \delta(k'-k) +\frac{r}{2} \delta(k+k') \right).
 \end{split}
 \end{equation}
Here, $\sqrt{2\pi}$ is imposed in order to fulfill the normalization condition. In addition, the delta functions are taken to be approximations where the modulations away from the main frequency $w_k = \hbar w k$ is neglected. This approximation can be viewed accurate because the scattering potential is an even and highly peaked function at origin, hence the wave-function is assumed to modulate in a narrow region around the boundary.\footnote{The corresponding real space wave function for this Fourier component is $\psi(x)= \frac{1+t}{2} e^{ikx}+ \frac{r}{2} e^{-ikx}$. Since the potential is even and highly peaked, one can show that our assumption corresponds to the idea that the wave function changes in a narrow region near the boundary.}
\par This approximation allows us to obtain the following relation for the integral in Eq. (\ref{eq:integral}):
\begin{equation}
\int_{-\infty}^{\infty}\diff x V^*(x) \psi(x) = \sqrt{\frac{\pi}{2}} (1+r+t) V^*(k).
\end{equation} 

This allows us to express the excitation coefficient as:
\begin{equation} \label{eq:excitation}
e_k = -\sqrt{\frac{\pi}{2}} (1+r+t)  \frac{V^*(k)}{(\Omega-E_k)},
\end{equation}

In order to solve Eq.s (\ref{eq:boundary2}) and ($\ref{eq:boundary3}$), we propose a novel approach called the ``single-frequency sampling method." This approach may be of significance for wide range of applications in information theory and signal processing; the details are presented in the next section. 

\subsection{Single Frequency Sampling Method with UV-Cutoff}

The sampling method we propose here allows us to treat functions, $V(x)$, which have a converging Fourier Transform, $V(k)$ with an ultraviolet cutoff. If the sufficient conditions of the Fourier Transform are not satisfied, it may not always be possible to interpret some functions in their conjugate domain. In the case of the TLS, the coupling coefficient $V(k)$ is an analytic function in k-space, however, its Fourier component $V(x)$ is not. One way to deal with this problem is to linearize the coupling coefficient $V(k)$ around the region of interest, e.g. the resonance of the atom in Ref. \cite{fan}. This leads us to obtain resulting equations that are correct in a region of convergence. However, this assumption, despite being highly informative regarding the behaviour of the system, may leave out some interesting properties, as in the case where the radius of convergence is too small. Here, we propose to abandon this assumption and use the sampling method to obtain more precise solutions. As we show below, this method also allows us to comment on the convergence of the linearization.\par 

We begin by focusing our attention to the solution of Eq.s (\ref{eq:boundary2}) and (\ref{eq:boundary3}). The following derivation is of significance as it allows us to sample desired signal out of an infinite source, which is the diverging scattering potential $V(x)$. We rewrite  Eq. (\ref{eq:boundary2}) in the following form: 

\scriptsize
\begin{align*}
0&=- i \hbar v_g \frac{\partial u_R (x)}{\partial x} e^{ikx} + e_k V(x) \\
&= - i \hbar v_g \frac{\partial u_R (x)}{\partial x} + \frac{1}{\sqrt{2\pi}} e_k \int_{-k_{max}}^{k_{max}} \diff k' e^{ik'x} V(k') e^{-ikx}  \\
&= - i \hbar v_g \int_{-\infty}^\infty \diff x \frac{\partial u_R (x)}{\partial x} +  \frac{1}{\sqrt{2\pi}} e_k   \int_{-\infty}^\infty \diff x \int_{-k_{max}}^{k_{max}} \diff k' e^{ik'x} V(k') e^{-ikx}  \\
&= - i \hbar v_g [u_R(\infty)-u_R(-\infty)] + \frac{1}{\sqrt{2\pi}} e_k  \int_{-k_{max}}^{k_{max}} \diff k' V(k') 2\pi \delta(k-k')  \\
&= - i \hbar v_g (t-1) +  \sqrt{2\pi} e_k V(k) 
\end{align*} \normalsize
where the term $e^{ikx}$ ``samples" the coupling coefficient between the two-level atom and a photon with wave vector $k$.\footnote{Here, one concern is whether one can switch the order of integration on the third row during this sampling process. This improper property is resolved with the introduction of the ultraviolet cutoff $k_{max}$. If the energy is allowed to be up to some finite energy scale, say Planck scale, then the order of the integration can be changed without any concern. }

After a similar sampling with $e^{-ikx}$, we obtain the following set of equations: 
\begin{subequations} \label{eq:sson}
\begin{align}
t \,i \hbar v_g -  i\hbar v_g  &= - 2\pi \frac{|V(k)|^2}{(\Omega-E_k)}  \frac{(1+r+t)}{2},  \\
r \, i \hbar v_g  &= -2\pi \frac{|V(k)|^2}{(\Omega-E_k)}  \frac{(1+r+t)}{2},
\end{align} 
\end{subequations}
where we obtain a linear set of equations with two unknown reflection and transmission coefficients, $r$ and $t$, respectively. 

This illustrates an application of our approach where we sample desired coupling coefficient, $V(k)$, for a photon with a wave-vector $k$, using multiplication with $e^{\pm ikx}$ and integrating. The term $e^{\pm ikx}$ behaves like a signal carrier between real ($x-$) and momentum ($k-$) phase spaces for the point $(x,k)$, where the integration over all real space ($x$) carries the signal from the real space onto the momentum space. With this method, the divergence of the real space representation of the signal becomes trivial, since the signal carrier $e^{ikx}$ samples only a portion of it, $V(k)$, which is convergent, which in turn justifies the usage of an ultraviolet cutoff. This methodology can be used for various applications in signal processing involving dual-phase spaces such as frequency-time and space-momentum.

\section{Further Characteristics}
In this section, we build on the theoretical foundations laid out in the previous section and discuss the implications of our approach such as the scattering coefficients, the S-matrix, spontaneous emission characterists and  a proof of concept using the dynamic electronic polarizability with additional remarks on the conditions under which the linearization assumption is valid. 

\begin{figure}
\centering
\includegraphics[width=9cm]{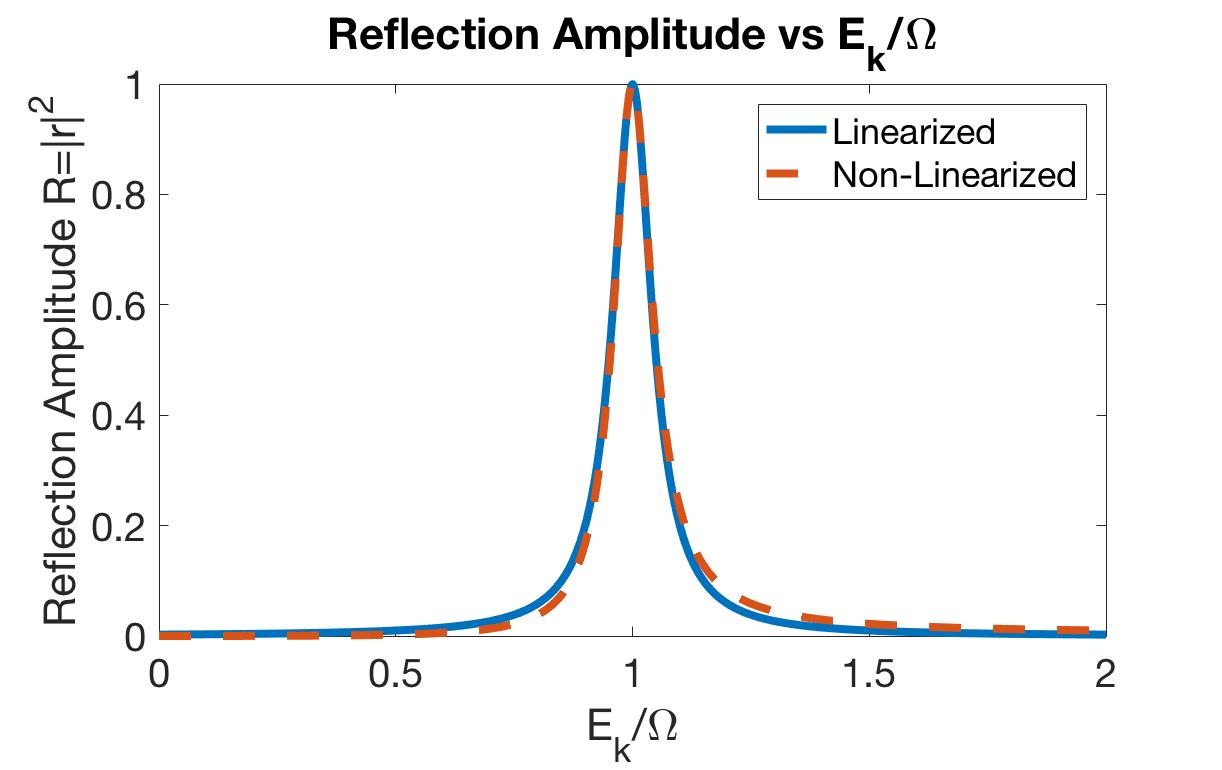} 
\caption{The reflection amplitudes for linearized and non-linearized coupling coefficients $V(k)$. $\xi = 0.05$.} \label{figtrans}
\end{figure}
\subsection{Scattering Coefficients and S-Matrix}

Once we solve the eigenvalue equation $\hat H \ket{E_k} = E_k \ket{E_k}$ and implement the sampling method, we are left with two linear equations which can be solved for two unknowns. After obtaining the transmission and reflection coefficients, we construct the S-matrix for this system including the non-radiative dissipation of the two-level atom. This S-matrix then can be used for future scattering experiments regarding two-level atoms.

We begin by solving the Eq. (\ref{eq:sson}) for transmission and reflection coefficients, t and r respectively, using straightforward algebra:

\begin{equation}  \label{eq:result}
\hfill 	t= \cos b e^{ib}, \quad \quad r=i\sin b e^{ib}.\hfill 
\end{equation} 
where the phase shift is given by $b= \arctan\left(\frac{\xi E_k }{(\Omega-E_k)} \right)$. Here, we define the dimensionless coupling parameter: 
\begin{equation}
\xi = \frac{d^2}{6 \epsilon_0 A \hbar v_g},
\end{equation} 
which can be obtained experimentally.
The reflection amplitude is given by: 
\begin{equation}
\begin{split}
R = |r|^2 = |i \sin b e^{ib}|^2 = \frac{\xi^2 E_k^2}{\xi^2 E_k^2 + (\Omega-E_k)^2}.
\end{split}
\end{equation} 
which is a modulated Lorentzian-like function, where the difference between half-reflection energies is given by $\Delta E_k /\Omega = \frac{2\xi}{1-\xi^2}$ The reflection amplitudes for linearized and non-linearized coupling coefficients $V(k)$ are sketched in Fig. 2.

The transfer matrix for a two-level atom with two different optical medium at the optical boundary is defined as: 
	\begin{equation*} 
\begin{pmatrix}
E_L^{(+)}  \\
E_L^{(-)} 
\end{pmatrix}
=
S_{[2\times2]}
\begin{pmatrix}
E_R^{(+)}  \\
E_R^{(-)}
\end{pmatrix}
\end{equation*} \normalsize
where $E_{L/R}^{(\pm)}$ is the amplitude of electric field with positive/negative wave-vector on the left/right of the two level atom.
Using the aforementioned transmission and reflection coefficients, the transfer matrix including dissipation is given by:

 \scriptsize
\begin{equation}  \label{eq:smatrix}
S_{2\times2}^{\text{two-level system}}=
\begin{bmatrix}
1 - i \xi \frac{E_k}{(\Omega - E_k)} & - i \xi \frac{E_k}{(\Omega - E_k)}\\
+ i \xi \frac{E_k}{(\Omega - E_k)} & 1 +i \xi \frac{E_k}{(\Omega - E_k)}
\end{bmatrix}
\end{equation} \normalsize 
 As can be seen from this transfer matrix, the coupling between the two-level atom and the photon is different for two energy levels equidistanced from resonance, where the coupling is higher for the higher energy level. This behaviour can be seen in Fig. \ref{figtrans}, where the reflection amplitude is modulated more significantly to the right of the resonance. This property of the transmission and reflection spectrum is a consequence of releasing the linearization assumption.

\subsection{Spontaneous Emission}
The spontaneous emission dynamics of a two-level atom inside a waveguide has been invastigated in Ref. \cite{Law}. However, it is important to note that even small changes in the decay rate can cause significant swings for the predicted lifetime of the two-level atom. Thus, we shall examine the decay rate of TLS, which is initially prepared in the excited state:
\begin{equation}
\ket{\alpha}=\ket{0,+}.
\end{equation}

One can project this state onto energy eigenstates given in Eq. (\ref{eq:eigenstate}) and then find the probability that the TLS remains in its excited state by taking into account the time evolution of the eigenstates as \cite{Law}:
\begin{equation}
P(t)=\left| \frac{1}{2\pi} \int \diff k |e_k|^2 e^{-iE_k t}  \right|^2.
\end{equation}

After straightforward algebra, we obtain the following expression for the probability of the two-level atom to remain in the excited state:
\begin{equation}
P(t)\simeq e^{-2 \frac{\xi \Omega}{(1+\xi^2)} t}.
\end{equation}

Comparing with the Ref. \cite{Law}, one can see that the spontaneous rate $\Gamma = 2 \frac{\xi \Omega}{(1+\xi^2)}$ is decreased by $\sim \frac{1}{\xi^2+1}$ if the linearization assumption is omitted. This stems from the fact that the non-linearized coupling coefficient is lower for the light-TLS interactions with energies $E_k < \Omega$. Therefore it is harder for the atom to decay, since it's coupling with the continuum is decreased.

\subsection{Proof of Concept: Dynamic Electronic Polarizability $a(w)$}
The dynamic electronic polarizability of a two-level atom has been studied extensively in Ref. \cite{polarizability} using Green's function method, whereas the scattering from a delta function shaped polarizability has been studied in Ref. \cite{photonics} using classical inhomogenous wave-equation. In our treatment so far, we included the real space scattering potential $V(x)$ as a function highly peaked at origin, almost as a delta function. Thus, combining the definition of the polarization vector $\vec{P}$ with our quantum mechanical treatment, it is possible to obtain the electronic polarizability $a(w)$ for the two-level atom. We start by citing the classical result obtained in Ref. \cite{photonics}:
\begin{equation}
\partial_x E_- - \partial_x E_+ = \frac{1}{\epsilon_0} k^2 \eta \alpha(w) E(x=0).
\end{equation} 
where $\eta$ is the particle number per unit area, which is $\frac{1}{A}$ in our configuration.

It is possible to bring the Eq. (\ref{eq:sson}) into the following form, which then can be compared to the above equation:
\begin{equation} \label{eq:one and only}
 i k \left(1-r- t \right) = \frac{1}{\epsilon_0}  k^2 \eta \frac{ \bar d^2}{ (\Omega-E_k)} \frac{(1+r+t)}{2}.
\end{equation} 

Considering the similarity, we denote the relation between the electric field and the polarization vector by $P= \eta \alpha(w) \delta(x) E$ and include the dissipation of the two-level atom \cite{fandissipation} to obtain the following dynamic electronic polarizability: 
\begin{equation}
\alpha(w) =  \frac{d^2}{3} \frac{1}{(\Omega-E_k- i \gamma/2)}
\end{equation} 
where $\bar d^2 = \frac{1}{3} d^2$ due to the averaging over all space \cite{average}. In our configuration, this result lays an evidence for the consistency of our method in classical and semi-classical limits, as long as rotating-wave approximation is valid. A more sophisticated discussion on this type of polarizability can be found in Ref. \cite{polarizability}.

\subsection{Remarks on Linearizations}
As a final remark, we emphasize the importance of the real space Hamiltonian given in Eq. (\ref{eq:dicke-Hamiltonian}). Omitting the linearization assumption in the coupling coefficient $V_k$ and solving the eigenvalue equation $\hat H \ket{E_k} = E_k \ket{E_k}$ results in the dynamic electronic polarizability $\alpha(w)$ in the interaction picture. This connection between the Green's function method in Ref. \cite{polarizability} and the classical version of the real space method in Ref. \cite{photonics} is obtained quantum mechanically through the real space version of Dicke--Hamiltonian given in Eq. (\ref{eq:dicke-Hamiltonian}). \par
Also, it is shown that the linearization of $V_k$, up to a multiplicative constant, does not pose any problem as long as $1 \pm \xi^2 \simeq 1$. This condition explains the success of the current literature on predicting the transport properties of this scattering experiment. 

Under the linearization assumption, we can represent the spontaneous decay rate as $\Gamma=2 \xi \Omega$, whereas the reflection and transmission amplitudes take the following form:
\begin{subequations}  
\begin{align}
 R=|r|^2 &= \frac{\Gamma^2}{(\Gamma+\gamma)^2+4(\Omega-E_k)^2},\\
  T=|t|^2 &= \frac{4(\Omega-E_k)^2+\gamma^2}{(\Gamma+\gamma)^2+4(\Omega-E_k)^2}.
\end{align}
\end{subequations}
where we can find that $T_{min}=\gamma^2/(\Gamma+\gamma)^2$ and $R_{max}=\Gamma^2/(\Gamma+\gamma)^2$. This result reveals that, in the resonance region, the 2-D materials can behave as two-level systems as the experimental findings of Ref. \cite{atachoca} suggests. We believe that our findings regarding TLS can explain further experimental results as long as the specimen can be approximated by a TLS and the coupling constant $\xi<<1$ is small. 

\section{Conclusions}
In this paper, we propose a novel approach that allows studying the interactions between the light and the TLS without making common assumptions in literature when the frequency $w$ of incoming photon is smaller than the cutoff frequency of the waveguide. In our approach, we use the coupling coefficient $V_k$ without linearization and study the conditions under which this approximation is valid. The sampling method we propose presents a novel tool for further analyzing a diverging real space scattering potential, $V(x)$, via a converging Fourier Transform, $V(k)$. \par 
Our studies show that for the two-level atom system, as long as the dimensionless coupling parameter $\xi<<1$ is small enough, the delta-scattering potential $V\delta(x)$, which is a consequence of the linearization, is a good model to describe this interaction. In order to show that the results obtained under our approach agrees with the existing literature, we  show that the dynamic electronic polarizability $a(w)$ is inline with the conventional atomic polarizability for a TLS.\par 
As a part of our future work, we plan to employ this approach to study the transport properties of a linear chain of atoms inside a waveguide. The assumption concerning the linearization of the coupling coefficient for a single atom remains valid for most or the transport properties. However, for multi-atoms cases, where the coupling plays a rather crucial role, the interaction between the light and atoms will require a more precise coupling coefficient calculations which may lead to different interesting transport properties. 

\section*{Acknowledgments} 
Both authors would like to thank B. Sankur for his insightful remarks and acknowledge generous resources provided by the Electrical \& Electronics Engineering Department at Bo\u{g}azi\c{c}i University.

\end{document}